\documentclass[aps,prl,twocolumn,amsmath,groupedaddress]{revtex4}

\def\al{\alpha}
\def\be{\beta}
\def\ga{\gamma}

\def\ep{\epsilon}

\def\et{\eta}

\def\ka{\kappa}
\def\la{\lambda}

\def\si{\sigma}

\def\ps{\psi}

\def\cl{{\cal L}}

\def\frac#1#2{\textstyle{{{#1} \over {#2}}}}
\def\pt#1{\phantom{#1}}
\def\prt{\partial}

\def\half{{\textstyle{1\over 2}}}
\def\lsim{\mathrel{\rlap{\lower4pt\hbox{\hskip1pt$\sim$}}
    \raise1pt\hbox{$<$}}}
\def\gsim{\mathrel{\rlap{\lower4pt\hbox{\hskip1pt$\sim$}}
    \raise1pt\hbox{$>$}}}

\def\etal {{\it et al.}}

\newcommand{\beq}{\begin{equation}}
\newcommand{\eeq}{\end{equation}}
\newcommand{\bea}{\begin{eqnarray}}
\newcommand{\eea}{\end{eqnarray}}
\newcommand{\bse}{\begin{subequations}}
\newcommand{\ese}{\end{subequations}}
\newcommand{\rf}[1]{(\ref{#1})}

\def\sqr#1#2{{\vcenter{\vbox{\hrule height.#2pt
         \hbox{\vrule width.#2pt height#1pt \kern#1pt
         \vrule width.#2pt}
         \hrule height.#2pt}}}}

\def \cep{\ep} 

\def\umnab{^\mu_{\pt{\mu}\nu\al\be}}
\def\mn{{\mu\nu}}

\def\lrprt#1{\stackrel{\leftrightarrow}{\prt_{#1}}}

\def\gm#1#2#3{g^{(M)}_{#1#2#3}}
\def\glA#1{g^{(A)}_{#1}} 
\def\gT#1{g^{(T)}_{#1}}
\def\tM#1#2#3{M_{#1#2#3}} 

\def\xx#1#2{\xi^{(#1)}_{#2}} 
\def\Tcoeff{(\xx 4 2 - 2 m_e \xx 5 8)}
\def\Acoeff{(\xx 4 4 - 2 m_e \xx 5 9)}
\def\Tcoef#1{(\xx 4 2 - 2 m_{#1} \xx 5 8)}
\def\Acoef#1{(\xx 4 4 - 2 m_{#1} \xx 5 9)}

\def\mgev{{\rm ~GeV}}
\def\mmm{{\rm ~m}^{-1}}

\def\tor#1#2#3{T^{#1}_{{\pt{#1}}#2#3}}

\begin{document}

\title{Constraints on Torsion from Lorentz Violation}

\author{V.\ Alan Kosteleck\'y$^1$, Neil Russell$^2$,
and Jay D.\ Tasson$^1$}
\affiliation{$^1$Physics Department, Indiana University, 
Bloomington, IN 47405, U.S.A.\\
$^2$Physics Department, Northern Michigan University,
Marquette, MI 49855, U.S.A.}

\date{IUHET 510, December 2007; accepted in Physical Review Letters}

\begin{abstract}
{\noindent
Exceptional sensitivity to spacetime torsion can be achieved
by searching for its couplings to fermions.
Recent experimental searches for Lorentz violation
are exploited to extract new constraints 
involving 19 of the 24 independent torsion components
down to levels of order $10^{-31}$ GeV.
}
\end{abstract}

\maketitle 

In Einstein's general relativity,
gravity is the curvature of spacetime
and energy-momentum density is its source.
Among the numerous alternative theories of gravity,
one popular class of models involves introducing
an additional warping of spacetime
called torsion
\cite{hhkn,shapiro,hammond}.
In many models,
the torsion has spin density as its source.  
Some scenarios allow torsion waves to propagate through spacetime,
in analogy with the travelling curvature waves
that form gravitational radiation in general relativity.
In the special class of `teleparallel' models,
the curvature of spacetime is itself 
determined in terms of the torsion.

Theories extending general relativity 
via torsion are widely regarded 
as experimentally challenging to test 
because the effects of torsion typically are minuscule.
Nature contains many sources of energy-momentum density
sufficient to curve spacetime,
such as stars and planets.
However,
sources of spin density strong enough
to produce torsion effects are difficult to identify or create. 
Typical limits on torsion in the literature 
involve dynamical properties,
being obtained from searches for spin-spin interactions
or for torsion-mass effects
\cite{cfbsv}.

In this work,
we discuss an alternative approach to
searching for torsion,
based on the little-appreciated fact
that background torsion 
violates effective local Lorentz invariance.
The key point is that 
nonzero torsion over a region of spacetime
establishes a preferred orientation 
for a freely falling observer,
which is the defining criterion for 
local Lorentz violation \cite{akgrav}.
Certain tests of Lorentz symmetry
can therefore be reinterpreted as torsion searches.
Related ideas have been suggested by 
L\"ammerzahl \cite{cl}
and Shapiro \cite{shapiro}.
Here,
we use the exquisite sensitivities 
recently achieved in Lorentz-violation searches 
to extract tight new constraints on torsion components,
including many previously unbounded 
in the literature.

The recent surge of interest in tests of relativity 
stems from the realization
that tiny violations of Lorentz symmetry
could emerge from attempts to unify the known forces
\cite{ks}
and from the development 
of a comprehensive description 
of Lorentz and CPT violation
in the context of realistic effective field theory,
called the Standard-Model Extension (SME)
\cite{ck}.
The SME categorizes Lorentz violations
by the mass dimension of the  
corresponding operator in the Lagrange density,
which offers a simple measure of their expected size
\cite{kp}.
The physical effects are controlled by
coefficients for Lorentz violation,
and many experiments have been performed
to measure them
\cite{krtables}.
This work shows these experiments
can be reinterpreted as searches for nonzero torsion.
We note in passing that similar analyses might be relevant 
to non-torsion effects in other alternative gravity theories.

For the present work,
we suppose that the physical gravitational field
is described by a theory predicting
a nonzero torsion field in the vicinity of the Earth,
and we seek model-independent constraints on the torsion
insofar as possible.
The specifics of the torsion field 
will depend on details of the theory
and on the nature of the sources.
To minimize model dependence,
we take advantage of the fact that 
the dominant effects of nonzero torsion 
include various modifications of particle behavior 
arising from couplings to the torsion as a background.
It is reasonable to approximate
the leading-order torsion background as constant
in a suitable reference frame relative to the torsion source.
Analysis of subleading terms with torsion derivatives
might conceivably yield additional results of interest,
but this issue lies outside our present scope.

The constant torsion background
establishes a preferred orientation in the specified frame,
producing effective local Lorentz violation.
The Lagrange density describing 
particle couplings to background torsion 
can therefore be matched to 
the SME Lagrange density describing 
particle behavior in the presence of Lorentz violation.
The match permits us to extract 
constraints on many torsion components 
from existing limits on SME coefficients for Lorentz violation. 
Note that this procedure is sufficiently powerful 
to provide limits on models 
with both torsion and Lorentz violation
and to distinguish the two in many cases \cite{akgrav,bk}.
However,
since at present no compelling evidence exists
for either Lorentz violation or torsion,
our analysis here interprets experimental results
entirely in terms of torsion.
In what follows, 
we summarize the calculations and results of this proposal. 
The conventions are those of Ref.\ \cite{akgrav}.

The spacetime of general relativity is a Riemann manifold,
determined by the Riemann curvature tensor 
$\widetilde R\umnab$. 
A spacetime with torsion is a Riemann-Cartan manifold,
and it is specified by the generalized Riemann tensor
$R\umnab$
and the torsion tensor
$\tor\al\mu\nu$.
The tensor $R\umnab$ 
can be expressed as the sum of 
$\widetilde R\umnab$ and terms involving $\tor\al\mu\nu$.
Gravitational effects are neglibible
for our analysis of laboratory experiments,
so we can assume $\widetilde R\umnab \approx 0$.
The tensor $R\umnab$ 
is then determined by the torsion
and is nonzero only if the torsion is nonzero. 
Our analysis and results apply to most torsion theories 
predicting nonzero laboratory effects.
Teleparallel models are exceptions, 
as they require $R\umnab$ to vanish
and $\widetilde R\umnab$ to be nonzero.

The torsion tensor 
obeys $\tor\al\mu\nu = - \tor \al\nu\mu$
and therefore has 24 independent components.
It can be expanded
as a sum involving three Lorentz-irreducible pieces,
\bea
T_{\al\mu\nu} &=&
\frac 1 3 (g_{\al\mu}T_\nu - g_{\al\nu}T_\mu)
-\cep_{\mn\al\be} A^\be
+M_{\al\mu\nu},
\nonumber\\
T_\mu 
&\equiv&
g^{\al\be} T_{\al\be\mu} ,
\quad
A^\mu 
\equiv 
\frac 1 6 \cep^{\al\be\ga\mu} T_{\al\be\ga},
\nonumber\\
M_{\al\mu\nu} &\equiv& 
\frac 1 3 (T_{\al\mu\nu} + T_{\mu\al\nu}+ T_\mu g_{\al\nu})
- \frac 1 3 (\mu \leftrightarrow \nu).
\label{irreps}
\eea
The mixed-symmetry irreducible piece $M_{\al\mu\nu}$
satisfies the eight identities
$g^{\al\be} M_{\al\be\mu} = 0$ and
$\cep^{\al\be\ga\mu} M_{\al\be\ga} =0$,
leaving 16 independent combinations.

Our analysis sets constraints on torsion components 
through torsion couplings to Standard-Model fields.
In most of the literature, 
the torsion is assumed to be minimally coupled 
through its appearance in covariant derivatives. 
However,
nonminimal couplings are also possible.
In this work,
we focus on signals of torsion 
arising from both minimal and nonminimal couplings to fermions.
Couplings to bosons,
including photons and gravity,
are discussed briefly below
but yield constraints of lesser interest
in the present context.

The general behavior of a fermion of mass $m$ 
in a Riemann-Cartan spacetime
with $\widetilde R\umnab \approx 0$
can be described by a hermitian Lagrange density 
with arbitrary torsion couplings.
We are interested in the constant-torsion approximation,
for which the arbitrary torsion couplings
can be replaced
with background solutions to the torsion field equations
taken as constant at leading order.
The corresponding effective Lagrange density
with all independent constant-torsion couplings 
of mass dimensions four and five can be written as 
\bea
\cl^T
&\supset&
\half i \bar \ps \ga^\mu \lrprt\mu \ps
- m \bar \ps \ps
+ \xi^{(4)}_1 T_\mu \bar\ps\ga^\mu\ps
\nonumber \\ &&
+ \xi^{(4)}_2 T_\mu \bar\ps\ga_5\ga^\mu\ps
+ \xi^{(4)}_3 A_\mu \bar\ps\ga^\mu\ps
+ \xi^{(4)}_4 A_\mu \bar\ps\ga_5\ga^\mu\ps
\nonumber \\ &&
+ \half i \xi^{(5)}_1 T^\mu \bar\ps\lrprt\mu \ps
+ \half \xi^{(5)}_2 T^\mu \bar\ps \ga_5 \lrprt\mu \ps
\nonumber \\ &&
+ \half i \xi^{(5)}_3 A^\mu \bar\ps\lrprt\mu \ps
+ \half \xi^{(5)}_4 A^\mu \bar\ps \ga_5\lrprt\mu\ps
\nonumber \\ &&
+ \half i \xi^{(5)}_5 {M^\la}_{\mn} \bar\ps \lrprt\la \si^\mn\ps
+ \half i \xi^{(5)}_6 T_\mu \bar\ps \lrprt\nu \si^\mn\ps
\nonumber \\ &&
+ \half i \xi^{(5)}_7 A_\mu \bar\ps \lrprt\nu \si^\mn\ps
+ \half i \xi^{(5)}_8 \cep^{\la\ka\mn} 
T_\la \bar\ps \lrprt\ka \si_\mn\ps
\nonumber \\ &&
+ \half i \xi^{(5)}_9 \cep^{\la\ka\mn} 
A_\la \bar\ps \lrprt\ka \si_\mn\ps .
\label{lag}
\eea
Here,
non-torsion couplings are disregarded,
and covariant derivatives have been expanded
and approximated systematically.
The values of the coupling constants $\xi^{(d)}_j$
depend on the torsion theory considered.
For example, 
the special case of minimal coupling is 
recovered for $\xi^{(4)}_4 = 3/4$
with other couplings zero.

Each torsion component in $\cl^T$ is constant,
so it no longer has the particle Lorentz transformation properties 
\cite{akgrav}
of the original torsion field.
The theory therefore contains effective Lorentz violation.
For example, 
$A_\mu$ now behaves as four scalars 
under particle Lorentz transformations.
Moreover,
since the number of indices 
on the irreducible torsion components is odd,
all the terms in Eq.\ \rf{lag} violate effective CPT symmetry.  
Laboratory experiments can therefore in principle 
discern different torsion signals for particles and antiparticles.

In $\cl^T$,
each constant torsion field 
and its associated coupling constant 
can be reinterpreted as a constant coefficient
for a fermion field operator 
having mass dimension three or four.
With this reinterpretation,
$\cl^T$ can be matched
to the Minkowski-spacetime limit
of the fermion sector of the minimal SME
\cite{akgrav},
which includes terms of dimension four or less.
In this sector,
the coefficients for Lorentz violation
controlling CPT-odd effects are conventionally denoted
$a_\mu$, $b_\mu$, $e_\mu$, $f_\mu$, 
and $g_{\mu\nu\al} = - g_{\nu\mu\al}$.
The latter can be decomposed into components
$\gT\mu$, $\glA\mu$, and $\gm \mu \nu \al$
in analogy with Eq.\ \rf{irreps}.
However,
in the present context
only the combination 
$b_\mu - m \glA\mu$ and the mixed irreducible component 
$\gm \mu\nu\al$
are independent observables at leading order
\cite{ck,akgrav,ba}.
Matching these combinations to irreducible torsion components gives
\bea
b_\mu - m \glA\mu &=& 
\hskip -1pt
- (\xi^{(4)}_2 - 2m \xi^{(5)}_8) T_\mu 
- (\xi^{(4)}_4 - 2m \xi^{(5)}_9) A_\mu ,
\nonumber \\
\gm \mu \nu \al &=& 
\hskip -1pt
- 2 \xi^{(5)}_5 \tM \al\mu\nu ,
\label{match}
\eea
which fixes the correspondence between 
observable effective torsion couplings 
and minimal SME coefficients 
involving field operators up to dimension four.
This correspondence could be extended to 
operators of arbitrary dimension 
using coefficients in the nonminimal SME.

To identify explicit constraints on torsion components,
the reference frame in which 
the leading-order torsion background is constant 
must be identified.
This frame depends on the underlying theory
and factors such as the source of torsion.
We first suppose that 
the constant-torsion approximation holds
everywhere within the solar system,
as might occur in models with 
torsion originating outside the solar system,
perhaps on galactic or cosmological scales.
It is then appropriate and convenient to adopt
the Sun-centered celestial-equatorial frame 
with cartesian coordinates $(T,X,Y,Z)$ 
that is widely used in reporting results 
of searches for Lorentz violation
\cite{mmbklr}.
In this frame,
the $Z$ axis is parallel to the Earth's rotation axis,
and the $X$ axis points towards the vernal equinox.

Laboratory measurements of 
$b_\mu - m \glA\mu$ and $\gm \mu \nu \al$
have been reported in the Sun-centered frame
for various fermion species
\cite{krtables}.
Since torsion is a geometric phenomenon,
its couplings can reasonably be assumed 
to be flavor independent. 
A search through the available measurements
with this flavor independence in mind reveals that
the sharpest sensitivities to torsion effects
emerge from Zeeman measurements with a dual maser 
\cite{cane}
and from studies of a spin-polarized torsion pendulum 
\cite{heckel}.

The results from the dual-maser experiment 
include six independent measurements
of combinations of coefficients for Lorentz violation
involving the neutron,
obtained by searching for modulations of the maser signal 
associated with the rotation of the Earth and
its revolution about the Sun.
Applying the match \rf{match},
we extract six constraints 
on combinations of torsion coefficients:
\bea
&&
| \Tcoef n T_X + \Acoef n A_X 
\nonumber \\ && 
\qquad\qquad
\qquad
+ 2 m_n \xx 5 5 \tM T Y Z |
< 1.6 \times 10^{-31}\mgev ,
\nonumber\\
&&
|\Tcoef n T_Y + \Acoef n A_Y 
\nonumber \\ && 
\qquad\qquad
\qquad
+  2 m_n \xx 5 5 \tM T Z X |
< 1.9 \times 10^{-31}\mgev ,
\nonumber\\
&&
| \cos \et [ \Tcoef n T_T + \Acoef n A_T 
\nonumber \\ && 
\quad\qquad
+ 2 m_n \xx 5 5 \tM Y Z X ]
\nonumber \\ && 
\quad
    + 2 m_n \xx 5 5 \sin \et (2 \tM T T X - \tM Y Y X) | 
\nonumber \\ && 
\qquad\qquad
\qquad\qquad
\qquad\qquad
< 2.0 \times 10^{-27}\mgev , 
\nonumber\\
&&
| 2 m_n \xx 5 5 (\tM T T Z + \tM X X Z) |
 < 3.6 \times 10^{-27}\mgev , 
\nonumber\\
&&
| \Tcoef n T_T + \Acoef n A_T 
\nonumber \\ && 
\qquad\qquad
\quad
+ 2 m_n \xx 5 5 \tM X Y Z |
 < 3.8 \times 10^{-27}\mgev , 
\nonumber\\
&&
| 2 m_n \xx 5 5 [\cos \et(2 \tM T T Z - \tM X X Z)
\nonumber \\ && 
\qquad
- \sin\et(\tM T T Y + \tM Z Z Y)] |
 < 1.6 \times 10^{-27}\mgev ,
\nonumber\\
\label{dualmaser}
\eea
where $m_n$ is the neutron mass 
and $\et\simeq 23.4^\circ$ is the inclination 
of the orbital plane of the Earth
relative to the $X$-$Y$ plane.
Ongoing dual-maser experiments can be expected 
to improve these constraints
in the future.

The experiment with a spin-polarized torsion pendulum
sought potential Lorentz-violating signals
modulated by the rotation of the Earth
and a laboratory rotation of the pendulum.
The results include measurements 
of three independent combinations 
of coefficients for Lorentz violation involving the electron
\cite{heckel}.
Using Eq.\ \rf{match},
we obtain three additional constraints
on torsion components:
\bea
&&
|\Tcoeff T_X + \Acoeff A_X 
\nonumber \\ && 
\qquad\qquad
\quad\qquad
+ 2m_e \xx 5 5 M_{T Y Z}|
< 4.8 \times 10^{-31} \mgev , 
\nonumber \\
&&
|\Tcoeff T_Y + \Acoeff A_Y 
\nonumber \\ && 
\qquad\qquad
\quad\qquad
+ 2m_e \xx 5 5 M_{T Z X}|
< 5.0 \times 10^{-31} \mgev , 
\nonumber \\
&&
|\Tcoeff T_Z + \Acoeff A_Z 
\nonumber \\ && 
\qquad\qquad
\quad\qquad
+ 2m_e \xx 5 5 M_{T X Y}|
< 7.8 \times 10^{-30} \mgev .
\nonumber \\ 
\label{torpend}
\eea
Here,
$m_e$ is the electron mass.
Further analysis of the torsion-pendulum data
allowing for modulations 
from the annual revolution of the Earth 
could yield additional constraints 
of type similar to the last few in Eq.\ \rf{dualmaser}. 

\begin{table}
\begin{tabular}{|c|c||c|c||c|}
\hline
&&&&\\[-9pt]
\mbox{Quantity}&{\mbox{Sensitivity}}&
\mbox{Quantity}&{\mbox{Sensitivity}}& 
{\mbox{Source}}\\ 
[3pt] 
\hline
&&&&\\[-9pt]
$\xx 4 2 T_T$&$10^{-27} \mgev$ &$\xx 5 8 T_T$&$10^{-27}$ & \\
$\xx 4 2 T_X$&$10^{-31} \mgev$ &$\xx 5 8 T_X$&$10^{-31}$ & S \\
$\xx 4 2 T_Y$&$10^{-31} \mgev$ &$\xx 5 8 T_Y$&$10^{-31}$ & S \\
$\xx 4 2 T_Z$&$10^{-29} \mgev$ &$\xx 5 8 T_Z$&$10^{-26}$ & S,E \\
[3pt]
$\xx 4 4 A_T$&$10^{-27} \mgev$ &$\xx 5 9 A_T$&$10^{-27}$ & \\
$\xx 4 4 A_X$&$10^{-31} \mgev$ &$\xx 5 9 A_X$&$10^{-31}$ & S \\
$\xx 4 4 A_Y$&$10^{-31} \mgev$ &$\xx 5 9 A_Y$&$10^{-31}$ & S \\
$\xx 4 4 A_Z$&$10^{-29} \mgev$ &$\xx 5 9 A_Z$&$10^{-26}$ & S,E \\
[3pt]
&&$\xx 55 \tM T T X$&$10^{-27}$ & \\
&&$\xx 55 \tM T T Y$&$10^{-27}$ & \\
&&$\xx 55 \tM T T Z$&$10^{-28}$ & \\
&&$\xx 55 \tM X X Z$&$10^{-27}$ & \\
&&$\xx 55 \tM Y Y X$&$10^{-27}$ & \\
&&$\xx 55 \tM Z Z Y$&$10^{-27}$ & \\
&&$\xx 55 \tM T X Y$&$10^{-26}$ & S,E \\
&&$\xx 55 \tM T Y Z$&$10^{-31}$ & S \\
&&$\xx 55 \tM T Z X$&$10^{-31}$ & S \\
&&$\xx 55 \tM X Y Z$&$10^{-27}$ & \\
&&$\xx 55 \tM Y Z X$&$10^{-27}$ & \\
[3pt]
\hline
\end{tabular}
\caption{
Sensitivities to torsion components 
associated with SME operators of dimension three 
(first pair of columns)
and dimension four (second pair).
Modulus signs are suppressed. 
The third column indicates sensitivity
to Sun-sourced (S) and Earth-sourced (E) torsion
(see text).
}
\end{table}

To gain some feeling for the scope of the sensitivities 
to galactic or cosmological torsion
achieved in these experiments,
it is useful to tabulate the results
assuming that only one torsion component
is nonvanishing at a time.
Table I summarizes the best sensitivities achieved
under this assumption.
The first pair of columns displays results 
to torsion couplings involving dimension three operators,
while the second pair of columns shows
results for the dimension-four case.
The table reveals that 19 of the 24 torsion components
are accessible to these two laboratory searches at this order.
The results for $T_\mu$ and $M_{\al\mu\nu}$ are firsts, 
while those for $A_\mu$ improve existing ones 
in magnitude and detail.  
The sensitivities attained are of potential interest 
for model building,
as a torsion magnitude of $10^{-27}$ GeV
is roughly comparable to that of the metric laplacian 
on the surface of the Earth.
If torsion is strictly minimally coupled,
then only the antisymmetric irreducible components $A_\mu$
are constrained,
and the best sensitivities achieved are 
\bea
|A_T| < 2.9 \times 10^{-27} \mgev 
\simeq 1.5 \times 10^{-11} \mmm ,
\nonumber\\
|A_X| < 2.1 \times 10^{-31} \mgev  
\simeq 1.1 \times 10^{-15} \mmm ,
\nonumber\\
|A_Y| < 2.5 \times 10^{-31} \mgev  
\simeq 1.3 \times 10^{-15} \mmm ,
\nonumber\\
|A_Z| < 1.0 \times 10^{-29} \mgev 
\simeq 5.3 \times 10^{-13} \mmm .
\eea

In some of the torsion literature,
the primary tensor is a linear combination
of torsion components known as the contortion tensor
$K_{\mu\al\nu} = - K_{\nu\al\mu}$.
The contortion also has 24 independent components 
and can be defined as 
\bea
K_{\mu\al\nu} &\equiv&
\half ( T_{\mu\al\nu} - T_{\nu\al\mu} + T_{\al\mu\nu}) 
\nonumber\\
&\equiv&
\frac 1 3 (g_{\mu\al}K^{(T)}_\nu - g_{\nu\al} K^{(T)}_\mu)
- \cep_{\mu\al\nu\be} K^{(A)\be} 
+ K^{(M)}_{\mu\al\nu}.
\nonumber\\
\label{cont}
\eea
The second equation displays the decomposition of the 
contortion into Lorentz-irreducible pieces,
which are related to the Lorentz-irreducible torsion components by
\bea
K^{(T)}_\nu = T_\nu , 
\quad
K^{(A)}_\nu &=& \half A_\nu , 
\quad
K^{(M)}_{\mu\al\nu} = M_{\al\mu\nu}.
\label{contirreps}
\eea
The constraints \rf{dualmaser}, \rf{torpend}
and the entries in Table I
therefore also yield immediate results for the contortion.

The effective Lagrange density \rf{lag}
is an expansion in only one fermion species.
Torsion couplings involving more than one fermion,
such as $T_\mu \bar\ps_1\ga^\mu\ps_2$,
could also be considered.
The corresponding SME observables involve particle flavor changes,
so relevant experiments would include meson oscillations
or neutrino oscillations.
However,
the sensitivities to torsion achieved in such experiments
are weaker than those displayed in Table I.

The sensitivities listed in Table I
are obtained from fermion-torsion couplings.
Other possibilities can also be considered.
In the photon sector,
the natural electromagnetic field strength  
$F_\mn$
is the exterior derivative of the four-vector potential,
and it has no torsion coupling.
If instead $F_\mn$ is defined
via the covariant derivative,
the resulting minimal torsion coupling
violates U(1) gauge invariance.
However,
nonminimal torsion couplings to photons
that preserve U(1) gauge invariance
can be introduced in an effective-field expansion
in powers of $F_\mn$.
In the approximation of constant torsion background,
the corresponding leading-order terms
lie in the nonminimal SME and are of dimension five,
so the resulting constraints 
are expected to be of lesser interest.
The best existing sensitivity to terms of this type 
comes from studies of birefringence in the
cosmic microwave background 
\cite{mewes07},
but this involves cosmological scales 
rather than laboratory ones.
Couplings to other gauge bosons
in the electroweak and strong sectors
also lead to sensitivities of lesser interest.
In the gravity sector,
measurements have recently been obtained
for some coefficients for Lorentz violation
\cite{gravexpt}
that may imply bounds on curvature-torsion couplings,
but these are also expected to be of lesser interest
in the present context.

The above results assume background torsion
on the scale of the solar system
with dominantly constant components
in the Sun-centered cartesian frame.
On phenomenological grounds one can consider 
instead `Sun-sourced' torsion,
having an approximate azimuthal symmetry 
centered on the Sun and 
with symmetry axis normal to the ecliptic plane.
In this scenario,
the torsion components in the corresponding coordinates
are approximately constant over the Earth's orbit.
As a result,
the revolution of the Earth about the Sun 
produces no modulation at first approximation,
so the relevant bounds on torsion components
become restricted to those extracted from
Lorentz-violation searches 
involving the rotation of the Earth
or a laboratory rotation of the apparatus.
The corresponding results 
include the first two dual-maser measurements
in Eq.\ \rf{dualmaser} 
and the three torsion-pendulum ones
in Eq.\ \rf{torpend}. 
The Sun-sourced torsion sensitivities attained
are those in Table I
marked with an S in the final column.

In a similar vein,
one can consider `Earth-sourced' torsion
having an approximate azimuthal symmetry 
centered on the Earth and with symmetry axis 
along the Earth's rotation axis.
In principle,
comparison of data between laboratories at different latitudes 
could reveal effects varying with distance from the rotation axis,
but at fixed latitude
the torsion components are approximately constant 
in the corresponding coordinate system.
This means neither the rotation nor the revolution of the Earth 
modulate torsion effects in a given laboratory,
so Lorentz-violation searches 
for sidereal and annual variations are irrelevant.
In effect,
in a given laboratory,
only measurements arising from rotations of the apparatus
or direct comparisons of the behavior
of particles and antiparticles
can reveal torsion couplings of this type,
so only the third constraint in Eq.\ \rf{torpend} applies.
The resulting sensitivities attained
are those marked with an E in the final column of Table I.

This work was supported in part
by DOE grant DE-FG02-91ER40661
and by NASA grant NAG3-2194.

\end{document}